\RequirePackage{colortbl}

\pdfoutput=1 
\documentclass{JINST}

\usepackage{makecell}
\usepackage{todonotes}

\title{Shielding Design for the ISODAR Neutrino Experiment}

\author{Adriana Bungau$^a$, 
Jose Alonso$^a$, Larry Bartoszek$^b$ $^c$,  Janet M. Conrad$^a$, Edward Dunton$^b$, Michael H.
Shaevitz$^b$\\
\llap{$^a$}Massachusetts Institute of Technology, Cambridge, MA 02139, USA\\
\llap{$^b$} Columbia University, New York, NY 10027, USA\\ 
\llap{$^c$}Bartoszek Engineering, 818 W. Downer Place, Aurora, IL 60506-4904, USA\\ \\
E-mail: \email{abungau@mit.edu}}

\abstract{The IsoDAR sterile-neutrino search requires a very high intensity neutrino source.  For IsoDAR, this high intensity is produced using the high neutron
  flux from a 60 MeV, 10 mA proton beam striking a beryllium target that
  floods a sleeve of highly-enriched Li-7.  Through neutron capture the Li-7 is transmuted to Li-8, which beta-decays giving the desired high neutrino flux for
  very-short baseline neutrino experiments. The target can be placed very close to can existing large neutrino detector, which is typically located deep underground to reduce backgrounds. With such a setup, it
  is necessary to design a shielding enclosure for the target to prevent neutrons
  from causing unacceptable activation of the rock walls close to the
  target. Various materials have been studied including steel to thermalize the high energy neutrons and two new types of concrete developed by Jefferson Laboratory, one very light with
shredded plastic aggregate, and the other one enriched with high quantities of
boron. The shielding is asymmetrical, having a larger thickness
towards the detector in order to suppress the neutron and
gamma background in the neutrino detector.  Simulation results for rock activation and for detector backgrounds are presented.

}


\begin{document}

\newpage
\section{General Considerations}
  
The shielding design for high energy facilities and particle accelerators 
has become a key aspect of radiation protection due to the deep
penetration of high energy neutrons. Protection against neutron
radiation is obtained by using appropriate thicknesses and proper types 
of shielding materials that will slow down the high energy neutrons in
the first stage and then absorb the slow neutrons in
the second stage to reduce the neutron radiation to acceptable
levels. The IsoDAR experiment, in combination with KamLAND detector in Japan~\cite{KamLAND}, will perform sensitive short-baseline neutrino
oscillation searches and electro-weak measurements associated with beyond-the-standard-model physics. The
requirements for shielding and radiation protection, according to
Japanese law, must be in conformance with IAEA recommendations~\cite{IAEA}. 
In accordance with these requirements, 
the rock activation of the cavern wall due to artificially produced radionuclides 
must not exceed 0.1 Bq/g. The radiation exposure will be maintained 
as low as reasonably achievable through shielding around the beam-dump neutrino source.
The effectiveness of the shielding will be actively monitored by
radiation instruments located in the control room and by frequent
area-surveys performed by health physics personnel. 
Additional shielding is also required to reduce the unwanted neutron and
gamma interactions in the KamLAND detector that could cause significant backgrounds for the physics measurements.

\noindent
This paper is organised as follows. Section 2 describes the conceptual
design of the IsoDAR system as outlined in the Conceptual
Design Report~\cite{CDR}~\cite{IsoDAR}. Shielding considerations
regarding the neutron flux limitations, rock analysis and shielding 
material choice and combinations are presented in Section 3. The Monte
Carlo simulations and validation with experimental data and MCNPX (Monte Carlo N-Particle eXtended) studies
follow in Chapter 4. Material performance and design optimization are presented in section 5. The simulation results of rock activation and
radionuclides produced are discussed in Section 6 while the results
for neutron and photon background estimates in KamLAND detector are presented in Section 7.

\section{The IsoDAR Experiment}\label{sec:aaa}

At present, the particle physics community is placing a high priority on investigating neutrino masses, oscillations, and mixings~\cite{TNSA, Panel}.
Although in
the 3-neutrino oscillation model the three mixing angles and masses associated with three standard neutrino flavors are relatively
well known, anomalous results have been observed at LSND~\cite{LSND},
MiniBooNE~\cite{MiniBooNE} and short-baseline reactor experiments~\cite{reactor}. 
These inconsistencies can be explained with the hypothesis of a  (3+N) 
sterile neutrino model in which there are three light neutrino mass 
states and N more massive sterile neutrino states.  
IsoDAR is a proposed neutrino experiment that is being developed to do a definitive search for sterile neutrinos in the $\sim 1$ eV$^2$ $\Delta$m$^2$ mass region.

The IsoDAR experimental program will produce a high intensity 
$\bar{\nu}_{e}$ beam from the $\beta^-$ decay of $^{8}$Li.
The $\bar{\nu_{e}}$ can interact in the detector via the inverse beta 
decay (IBD) process: $\bar{\nu}_{e}$ + $p$ $\rightarrow$ $e^{+}$ + $n$.
In addition, IsoDAR also represents an important technological step, in terms of producing 
high-power cyclotrons that can be used for a number of physics and non-physics 
applications ~\cite{Daedalus},~\cite{cyclotron},~\cite{SAbe}.

\subsection{The  IsoDAR Design}\label{sec:ccc}

\begin{figure}[h!] 
\centering
\includegraphics[width=0.8\textwidth]{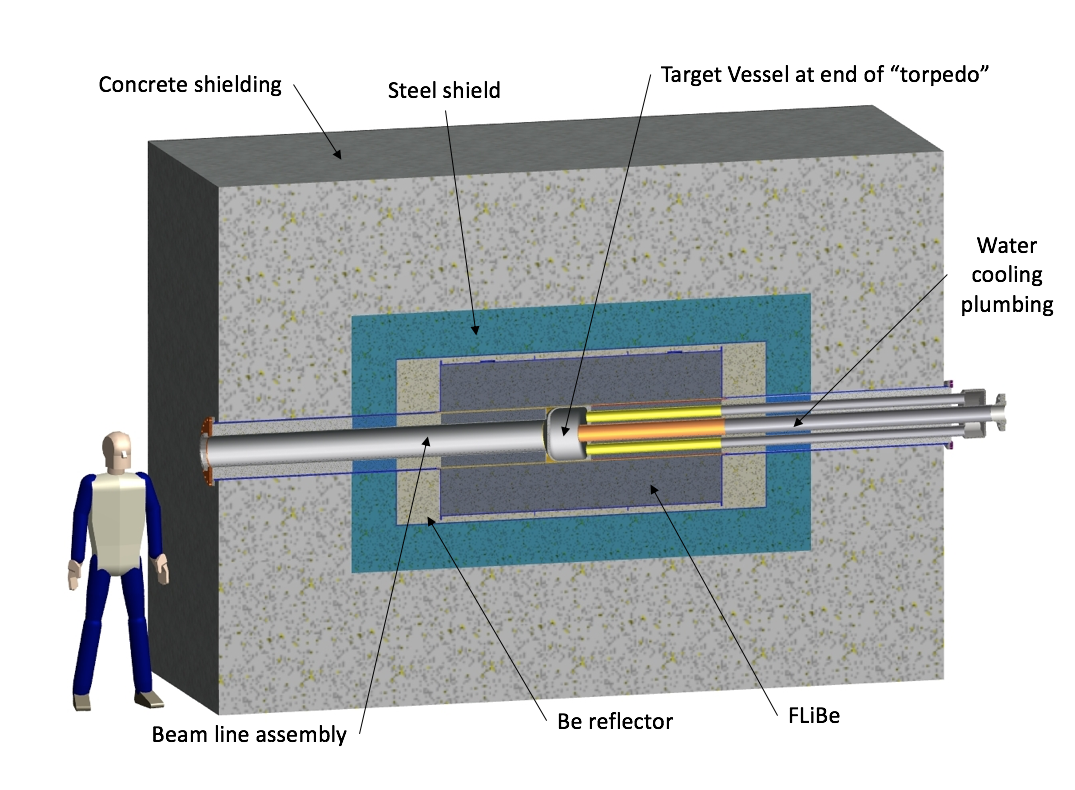}
\caption{The current layout of the target and shielding system. The target torpedo
is surrounded by LiBe sleeve (dark grey) and by graphite reflector
(grey). The target system is enclosed in shielding (steel and concrete
mixtures).}
\label{fig:1}
\end{figure}

The IsoDAR conceptual design consists of an ion source 
which injects up to 50 mA of H2+ into a high-power cyclotron that is 
required to capture and accelerate up to 5 mA of H2+ ions up to 60 MeV/ amu. Immediately 
after extraction, the molecular H2+ is dissociated into two protons, with
10 mA of protons delivered to the target system. The $^{9}$Be  target will 
produce a high neutron flux that  will then enter a sleeve surrounding 
the target, which contains 99.995\% pure $^{7}$Li (Fig.~\ref{fig:1}). The 
neutron capture on $^{7}$Li will create $^{8}$Li isotopes which
will then beta decay producing the electron antineutrinos. In the 
conceptual design the sleeve material is a mixture of lithium-floride 
and beryllium-floride (FLiBe), but recent studies have shown that a 
homogeneous mixture of lithium and beryllium with an optimum 
beryllium fraction mass of 75\% has a better efficiency of production of $^{8}$Li yield~\cite{sleeve}. The system is enclosed in a graphite 
reflector to enhance the neutron capture on $^{7}$Li. When coupled 
with the KamLAND detector, IsoDAR will observe 8.2 $\times$ 
$10^{5}$ reconstructed IBD events in five years of 90\% duty factor running. With this data set,
IsoDAR will provide 5$\sigma$ sensitivity to sterile neutrino oscillation
models as well as allow precision measurements of $\bar{\nu_{e}}-e$ scattering
and searches for the production and decay of exotic particles~\cite{report}.

\subsection{The IsoDAR Shielding }\label{sec:bbb}

The target assembly is to be placed in the Kamioka mine next to the 1 kton KamLAND liquid scintillator neutrino detector in one of the current utility areas near the detector. The space in this area will have approximate cross section 
dimensions of 2.25 m floor to ceiling and 3.5 m side to side. The 
remaining space after placing the target system will be used for 
shielding.  Preliminary calculations indicated that this space
will need to be enlarged, particularly in the vertical direction.
Minor excavation of the mine is allowed if it can be accomplished
without blasting. At the front and back end of the target system,
extra shielding material can be added if required.

\section{Shielding Considerations}\label{sec:xxx}

In the shielding design two types of radiation need to be
considered: neutrons and gamma rays. Most materials can attenuate
gamma rays as the thickness is increased with the higher the atomic
number and higher density of material giving greater the attenuation. The total neutron flux that is produced in the target system during the 
experiment depends on the target and sleeve geometries and
their materials.  A large fraction of these neutrons will be low energy
neutrons. Other neutrons reaching the cavern walls could also come from beam loss in the cyclotron and transport line to the target, but this can be minimized by reducing beam losses and providing active
monitors that interrupt the beam when abnormally high neutron levels
are detected. Additional shielding will be used to reduce the irradiation from the flux produced from unavoidable beam loss.

\subsection{Neutron Flux Limitation}\label{sec:ttt}

The neutrons that escape the shielding around the target will produce rock
activation in the cavern walls.  Radionuclides with half-lives shorter
than days, or even months, are of no consequence, but the progenitors 
of long-lived products (i.e. $^{60}Co$, $^{152}Eu$, $^{154}Eu$, $^{134}Cs$ etc.) need 
to be assessed. Rock samples were collected from various sites in the
mine in close proximity to the experiment location, and were irradiated 
in the reactor at MIT with a flux of $10^{18} n/cm^{2}$. An analysis of the irradiated
samples was performed at MIT and at LBNL and determined the presence of
cobalt and europium at the parts-per-million levels. As progenitors of
$^{152}Eu$, $^{154}Eu$ and $^{60}Co$ these concentrations provide a
measure of the upper limit of allowed neutron flux to exit the outer
surface of the shielding. From these measurements and calculations, a limit is set at $10^{-13}n/p/mm^{2}$ neutrons into the cavern walls.

\subsection{Shielding Materials}\label{sec:ooo}

The KamLAND detector is a delicate and sensitive large detection instrument and any mining activity in the close proximity of the detector needs to be minimized.  Therefore, minimizing the total volume of rock that must be removed for the IsoDAR target cavern is a prime requirement, which can be accomplished with the careful selection of target shielding material.  The choice of shielding is strongly dependent on neutron energy,
so an efficient combination of materials must shield against
the entire range of neutron energies. There are several factors that
must be taken into account when selecting the shielding materials. 
Considerations such as effectiveness, strength, resistance to damage 
and cost efficiency can affect radiation protection in many ways. 
While metals are strong and resistant to radiation damage, they 
undergo changes in their mechanical properties and degrade in time 
from radiation exposure. On the other hand, concrete materials are 
strong, durable and cost effective however they are weaker at elevated 
temperatures and less effective at blocking neutrons.
The materials used in this study were selected from 
a set of compound materials taking into account radiation shielding 
performance, physical and thermal properties and cost. 

First, the fast neutrons must be slowed 
down via inelastic scattering by using appropriate metal based attenuation 
materials like iron (carbon steel or stainless steel). Once the neutrons
are slowed down to thermal energies by inelastic collisions, then in a 
second stage the thermal neutrons are captured by the absorbing material.
Low-Z materials containing a high fraction of hydrogen (for example
water, plastic, concrete) can provide good neutron energy attenuation 
as a result of elastic scattering of neutrons on protons. At these
neutron energies the interaction cross section is high and the energy 
lost in a collision is significant. However, water is not a reliable
candidate as it can evaporate and leak while plastic materials are
expensive. Concrete is a good candidate as it is inexpensive and 
combines many of the good aspects required for shielding, 
particularly when different materials can be added as aggregates 
to the mixture. The Jefferson Laboratory (JLab) recently developed 
new shielding materials, a plastic concrete which performs better than
other materials for neutron thermalisation and a boron rich concrete which 
absorbs neutrons using less material~\cite{jlab_shielding}.  

\noindent By adding shredded plastic which
contains more hydrogen atoms to the concrete, one can increase its
ability to thermalise neutrons while decreasing its weight. Also, by
removing the grit and rocks that are normally found in concrete to
make it even lighter, the plastic concrete is basically two thirds of
the weight of the normal concrete and four times better at
thermalising neutrons. The boron rich concrete is 
basically Portland cement in which the normal rock/sand aggregate 
is replaced by pelletised boron carbide. In concrete, the neutrons are 
thermalised when they strike hydrogen atoms in the water molecules 
that are trapped during the concrete mixing process. Boron has a 
high neutron capture cross section and has been generally used in
for neutrino shielding in addition to concrete. The final product is a much 
better absorber and has the same consistency as ordinary concrete and
several layers of boron concrete were used in our studies.  In summary, 
the current IsoDAR shielding consists of a combination of high-Z and low-Z
materials which satisfies the requirements mentioned above.

\section{Monte Carlo Simulations}\label{sec:eee}

\subsection{The Physics Model}\label{sec:kkk}

There are several Monte Carlo simulation packages used for shielding
calculations and, for the calculations of the efficiency and performance of the
shielding materials described in this paper, the Geant4 simulation 
program was used \footnote{More information on the Geant4 physics models 
  can be found in the Physics Reference Manual:
  http://cern.ch/geant4/UserDocumentation/UsersGuides/
  PhysicsReferenceManual/html/PhysicsReferenceManual.html.}. 
The Geant4 modelling included the 
geometrical setup described before and the corresponding material 
properties, as well as the characteristics of the incoming proton
beam.

The physics package \textit{particle\_hp} was used in the current 
shielding simulations, which comprises a set of hadronic models
for proton and neutron inelastic interactions for an energy range up
to 200 MeV. The package works with Geant4 versions \textit{geant4.9.5} and 
later and uses evaluated nuclear data bases for inelastic interactions
of proton, neutron, deuteron, triton, He3, alpha and gamma.
\textit{Particle\_hp}  includes three physics lists: one for protons 
\textit{QGSP\_BIC\_PHP}, one for neutrons \textit{QGSP\_BIC\_NHP}, 
and one for all particles, i.e. neutrons, protons, deuterons, tritons, 
He3, alpha and gamma, \textit{QGSP\_BIC\_AllHP}. 
The abbreviation QGSP stands for the Quark Gluon String Parton model,
BIC for the Binary Intranuclear Cascade and HP for the high-precision neutron
package which includes evaluated neutron data for neutron interactions
below 20 MeV. The last two physics lists give the same results
for neutrons as the physics list  \textit{QGSP\_BIC\_HP}.  
The evaluated nuclear data libraries differ and, thus, the results of
the Monte Carlo simulations will depend on the library. Two databases 
\textit{ENDF/B-VII.1}~\cite{endf} and \textit{TENDL-2014}~\cite{tendl2014} 
were used for cross sections of primary and secondary particle
interactions. The \textit{ENDF/B-VII.1} library uses experimental data 
for projectile energies up to 150 MeV, which are essentially
nuclear reaction cross sections together with the distribution in 
energy and angle of the secondary reaction products. Also, it contains 
data only for 49 isotopes, including Be. The \textit{TENDL-2014} 
library uses some experimental data and \textit{TALYS}~\cite{talys}
calculations for projectile energies up to 200 MeV. It contains 
information for all isotopes and can be applied to all target 
materials but the best results are obtained for targets with
atomic number in the range 12-289.
For neutron energies below 20 MeV, the high-precision model
is employed, which uses \textit{ENDF/B-VII.1}, JENDL ~\cite{jendl},
MENDL-2 ~\cite{mendl} and other data libraries ~\cite{libraries}. The
Binary Intranuclear Cascade model is called for neutron energies above
20 MeV. This model includes a low energy nuclear de-excitation
model called the G4Precompound model which is called by the simulation
when the particle energy is below 100 MeV and when the nuclear
structure effects start to play an important role.

\subsection{Validation Studies}\label{sec:iii}

Differential neutron yields for several angles and various beam 
energies were measured in References~\cite{Waterman}-\cite{Harrison}.
New measurements of the neutron yield produced by a 62 MeV
proton beam on a thick beryllium target were
performed at Laboratori Nazionali del Sud (LNS) of INFN
using the existing superconducting cyclotron~\cite{Osipenko}. 
A 62 MeV proton beam with an operating beam current of 30-50 pA was
extracted from the cyclotron and transported through the beam
transport system to the target. The beryllium target 
had a thickness of 3 cm and 3.5 cm diameter. This thickness was
chosen to ensure complete absorption of the protons. The neutrons 
produced in the target were measured by the time-of-flight technique. 
Eight neutron detectors were installed around the target, at the same 
height with respect to the beamline, with different angles and with two 
different distances (150 cm and 75 cm). The electric charge deposited 
by the beam on the target was measured by a digital current integrator 
and used for absolute normalization of the data.

\begin{figure}[!h]
   \centering
   \includegraphics*[height=142mm, width=152mm]{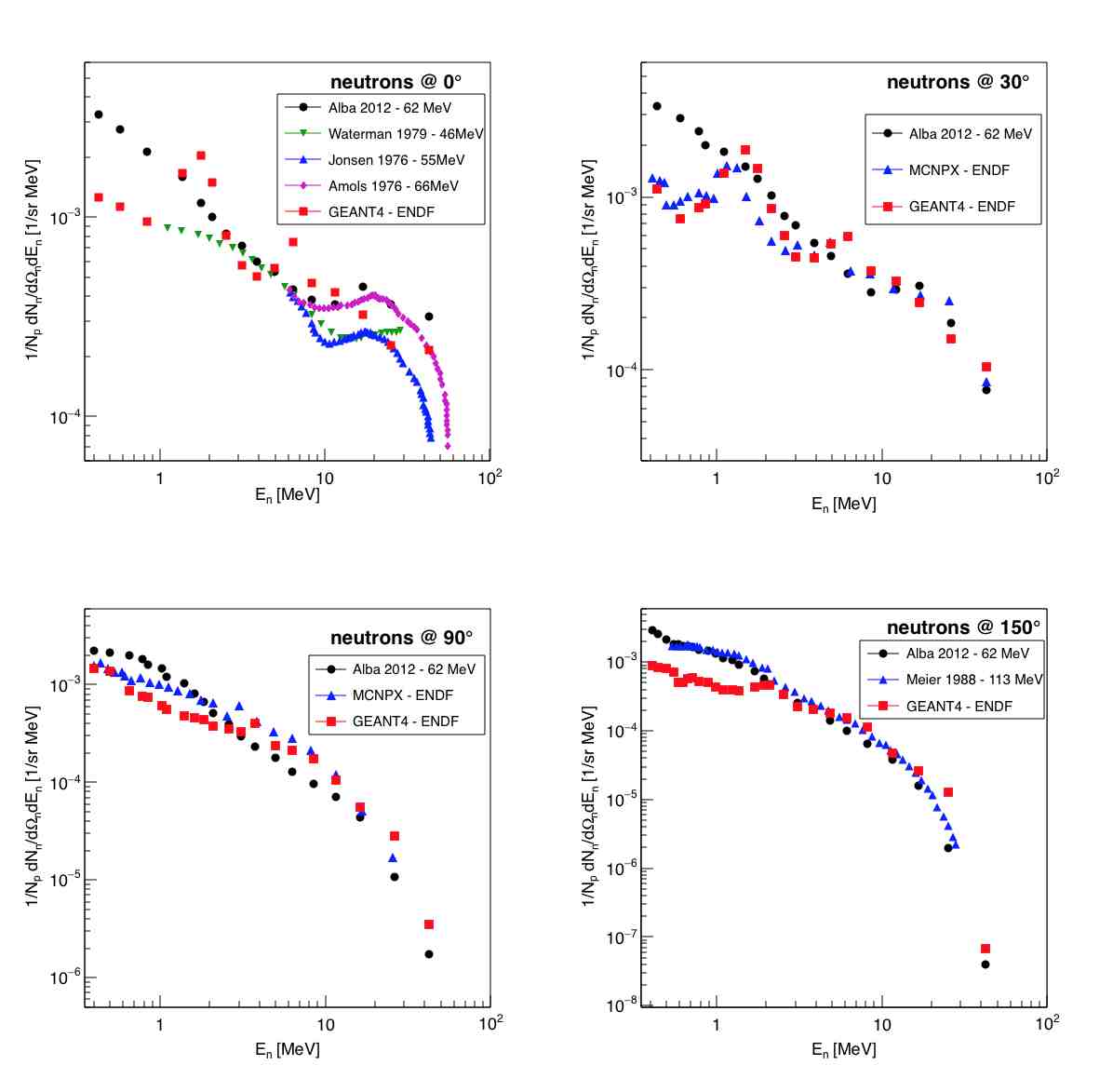} 
   \caption{Comparison of the \textit{particle\_hp} physics package predictions with
     the experimental data taken from this experiment, Ref. [2-5] and
     MCNPX simulations. The neutron yield was measured at
     $0^{\circ}$, $30^{\circ}$,  $90^{\circ}$ and  $150^{\circ}$. The
     proton energy is 62 MeV in all the Geant4 simulations.}
   \label{fig:data2}
\end{figure}

The target and the detector set up were modelled in the Geant4 simulations. The \textit{particle\_hp} physics package was used to simulate the neutron yield produced by 62 MeV protons and Fig.~\ref{fig:data2} shows the results at $0^{\circ}$, $30^{\circ}$,  $90^{\circ}$ and $150^{\circ}$. The comparison of the simulation with the experimental data taken from this experiment and also from Refs.~\cite{Waterman}, ~\cite{Johnsen}, and ~\cite{Almos} at $0^{\circ}$ for lower beam energies shows a significant disagreement in the lower neutron energy range. This disagreement
demonstrates lower beam energy data cannot be extrapolated to higher beam energies simply by an overall factor since the kinematic limits have to be taken into account. For low energy neutrons (below 10 MeV) there is a disagreement between data and simulation especially at lower angles. The \textit{ENDF VII} database library for particle cross sections was released in December 2011 and therefore it does not contain the experimental results measured in Ref.~\cite{Osipenko} at the later date. However, at larger angles, there is a good agreement between these two, even for low energy neutrons (above 2 MeV). 
At larger angles ($150^{\circ}$) there is a good agreement between the measured data in this experiment and Ref.~\cite{Meier} with the Geant4 code predictions for neutron energies above 2 MeV. At neutron energies below 1-2 MeV, the Geant4 predictions lie below the data. However these low energy neutrons will not make it through the shielding and therefore will not pose a problem for the rock activation studies.  

The Geant4 simulations were then compared with MCNPX results~\cite{mcnpx} using the ENDF data library for the measured cross sections. The comparison shows good agreement between the predictions of the two codes at larger angles, $30^{\circ}$ and $90^{\circ}$. There is a good agreement between the two codes and the data in some kinematic regions at larger angles and larger neutron energies.

The experimental measurements in ref.~\cite{Osipenko} which are not included in the  ENDF/B-VII proton database, show a difference about a factor of two greater than the values obtained with the two simulation packages below 1-2 MeV. However simulation values are higher in the range 2-70 MeV, so overall we expect the experimental and simulated neutron yields values integrated over the entire energy range (1-70 MeV), to be similar. This has been shown to be indeed the case in the validation study~\cite{sleeve} performed against the experimental measurements of neutron fluxes published in Ref.~\cite{Tilquin}.

As more experimental data for protons on beryllium become available, they will be added to the proton ENDF database increasing the accuracy of the model predictions. 
The current predictions of the 
\textit{particle\_hp} model rely on the existing tabulated
experimental data and will improve as the new data is implemented. 
The validation studies shown above as well as previous studies of 60
MeV protons on beryllium~\cite{sleeve} have shown that the 
\textit{particle\_hp} model with the ENDF data library describes 
the proton  inelastic interactions on Be  for 60 MeV incident energy 
better than any other theoretical model available justifying its selection for these studies.

\section{Shielding Design}\label{sec:ppp}

\subsection{Material Performance}\label{sec:ooo}

\noindent
Simulations were carried out to asses the performance of the
selected shielding materials. Neutron shielding materials with
variable thickness placed in different arrangements were examined 
and the comparisons are shown in this section. 
As the area where the target system and
the surrounding shielding will be located was once a construction tunnel 
for KamLAND, its dimensions are not overly generous. The cross
section of the cavern which measures roughly 2.3 m high by 3.5 m wide
leaves $\approx$50 cm from the target system to the ceiling. As the
available vertical space is the critical dimension in our Monte Carlo modelling, 
the figure of merit in the simulations is the neutron 
flux recorded on the ceiling, at 90 degrees with respect to the beam
direction. Initial studies using inner layers of plastic concrete for neutron 
moderation and outer layers of boron rich concrete for neutron 
absorption for the available $\sim$50 cm showed that the neutron 
flux was several orders of magnitude higher than the desired
value of $10^{-13}n/p/mm^{2}$. These results implied that rock 
will need to be removed to place an adequately shielded target in this tunnel. A summary of all the materials combinations, total shielding thickness and the neutron flux obtained in the simulations is shown in Table~\ref{tab:materials}.

 \begin{table}[h!]
\centering
\caption{\label{tab:materials} The materials combinations and the total shielding thickness used in the simulations.}
\vspace{0.2cm}
\begin{tabular}{| c | c | c |}
\hline\hline
 \thead{Materials Combinations} &  \thead{Total thickness} & \thead{Neutron flux }\\
\hline
\makecell{ plastic concrete and \\boron loaded concrete} & 100 cm & above $\Phi = 10^{-13} n/p/mm^{2}$ \\ 
\hline
\makecell {20 cm steel, plastic concrete\\ and boron loaded concrete} & 120 cm & some below $\Phi = 10^{-13} n/p/mm^{2}$\\
\hline
steel and boron loaded concrete & 120 cm & some below $\Phi = 10^{-13} n/p/mm^{2}$\\
\hline
steel and boron loaded concrete & 200 cm &below $\Phi = 10^{-13} n/p/mm^{2}$\\
\hline\hline
 \end{tabular}
 \end{table}

\begin{figure}[h!] 
\centering
\includegraphics[width=0.55\textwidth]{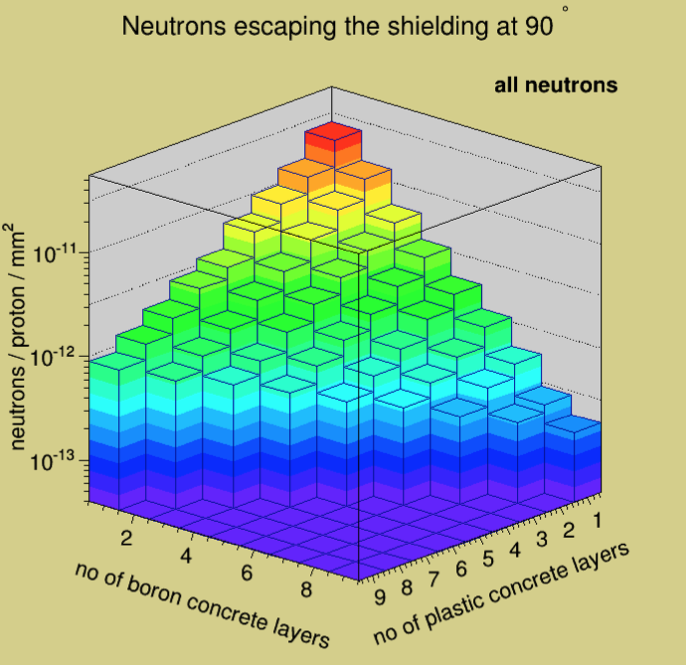}
\caption{The neutron flux escaping the shielding at a 90 degrees angle with respect to the proton beam direction. The shielding consists of combinations of layers of plastic concrete and boron rich concrete, each layer being 10 cm thick. The total shielding thickness is up to 100 cm. The neutron flux is above the desired value $10^{-13} n/p/mm^{2}$ for all combinations. (The purple combinations are not considered since they have total thicknesses greater than 100 cm.)}
\label{fig:002}
\end{figure}

\begin{figure}[h!] 
\centering
\includegraphics[width=1.0\linewidth]{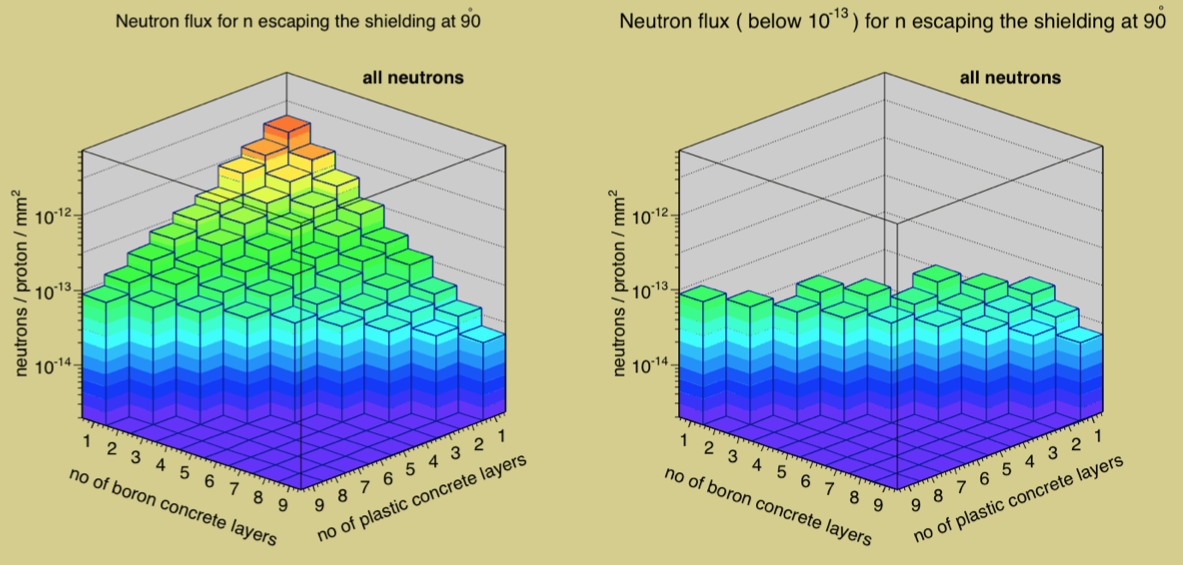}
\caption{The neutron flux out of the shielding at 90 degrees with respect to the beam direction. The shielding consists of two inner layers of steel of 20 cm total thickness (not shown), followed by combinations of layers of plastic concrete and boron rich concrete (left plot). Each layer has 10 cm thickness and the total shielding thickness is up to 120 cm. The right plot shows only those combinations for which the neutron flux is below the desired value $10^{-13} n/p/mm^{2}$.}
\label{fig:003}
\end{figure}

Assuming that a minimum of 50 cm of rock will
be removed from the cavern ceiling, a target shielding of total 
thickness of maximum 100 cm of various combinations of plastic concrete and 
boron rich concrete was considered.  The thickness of each material
layer was 10 cm in our simulations. The neutron flux recorded at 
90 degrees on a detector sphere was still above the desired value, 
$10^{-13} n/p/mm^{2}$ (Fig.~\ref{fig:002}). This was due to the fact
that the fast neutrons would escape the 100 cm of shielding and 
suggesting thus that more high-Z material is required to attenuate them. 
The inelastic scattering on high-Z atoms will reduce the neutron 
energy to a much lower  value such that they will be absorbed in the 
boron rich concrete layers. Therefore, in addition to the 100 cm shielding of plastic
concrete and boron rich concrete, two layers of 20 cm total thickness 
of steel were added. In this configuration, the inner layers are steel, followed by 
variable thicknesses of plastic concrete. The outer layers are boron
rich concrete to absorb the neutrons. The results are shown in 
Fig.~\ref{fig:003}.

\vspace{0.2 cm}

\begin{figure}[h!] 
\centering
\includegraphics[width=1.0\textwidth]{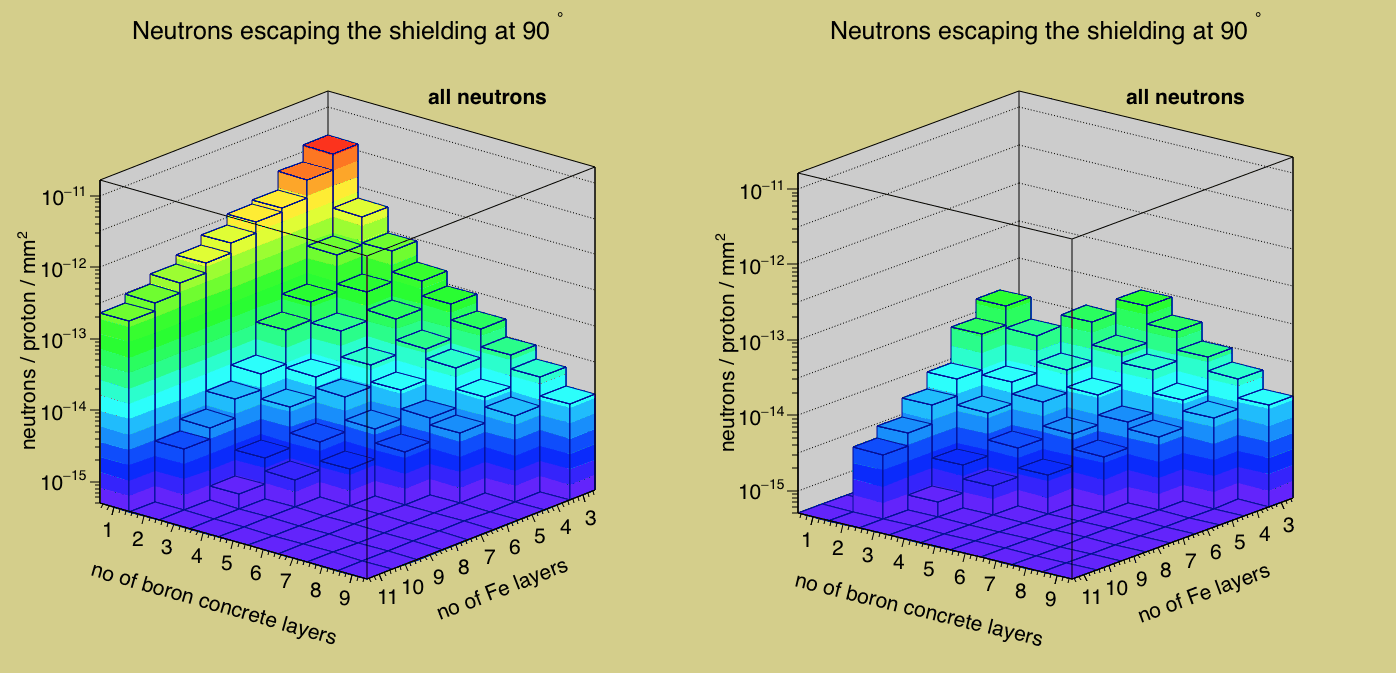}
\caption{The neutron flux escaping the shielding at 90 degrees with respect to the beam direction, using only combinations of layers of steel and boron concrete, but no plastic concrete. Each layer is 10 cm thick and the total thickness is up to 120 cm (left plot). For some combinations, the flux is much lower in this case than 
when plastic concrete was present and lower than 
the desired
  value $10^{-13} n/p/mm^{2}$ (right plot).}
\label{fig:004}
\end{figure}

There are several material combinations for which the neutron flux 
is lower than $10^{-13}n/p/mm^{2}$ (left plot). 
The flux is lowest for the combination 20 cm steel, 10 cm plastic concrete and 
90 cm boron rich concrete. 
As Fig.~\ref{fig:003} suggests, the optimum solution 
is for a minimum thickness of plastic concrete and indicates that
better results can be obtained for combinations of steel and boron
rich concrete only. The same total shielding thickness of 120 cm was
maintained but plastic concrete was removed
completely. The results are shown in Fig.~\ref{fig:004}. The neutron 
flux can be brought down to $10^{-15}n/p/mm^{2}$ for 90 cm of
steel and 30 cm of boron concrete. Because of the cost and total shielding 
weight, a final baseline configuration of 30 cm of steel and 90 cm of boron rich concrete was selected. The total mass of the baseline target system and 
shielding from this study comes to 165,331 kg.

\vspace{0.2 cm}

\vspace{0.2 cm}

\begin{figure}[th!] 
\centering
\includegraphics[width=0.85\textwidth]{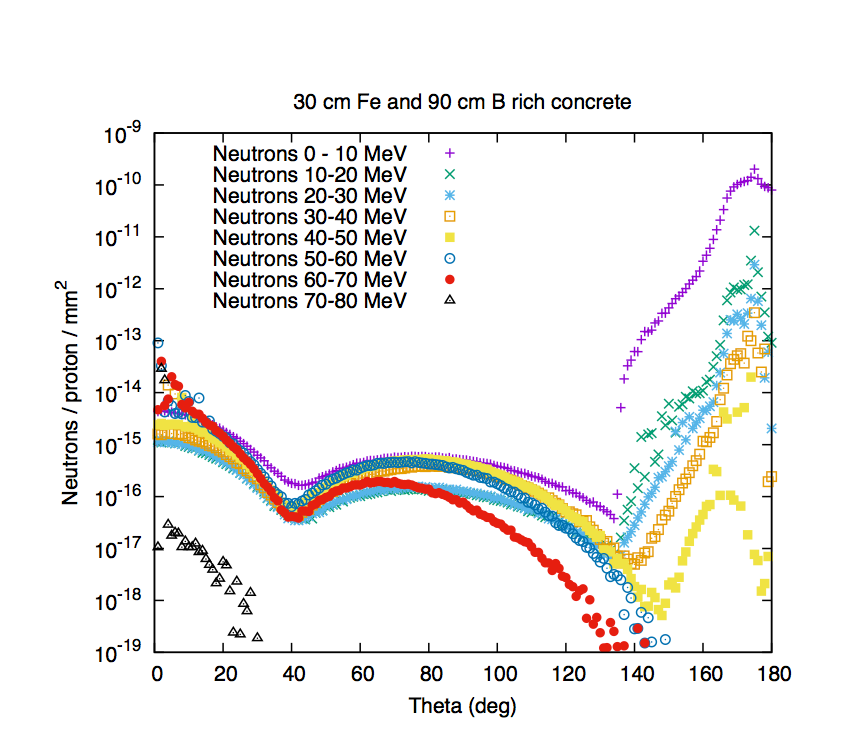}
\caption{The neutron flux detected on a sphere surrounding the
  shielding. The sphere radius is 350 cm. The shielding material
  consists of 30 cm Fe and 90 cm boron rich concrete. The high values
  of neutron flux at 0 and 180 degrees correspond to holes in the shielding for beam
  entry and target servicing. }
\label{fig:005}
\end{figure}

The neutron flux for this particular shielding configuration (30 cm steel and 90
cm boron rich concrete) was recorded on a detector sphere of radius 3.5 m and the results are shown in 
Fig.~\ref{fig:005}. The beam travels from right to left in Fig.~\ref{fig:005}. 
The lower flux values at 40 and 140 degrees correspond to the corners 
of the concrete shielding block and the higher values of flux above
140 degrees correspond to neutrons escaping into the space left for 
the wobbler magnets in front of the target where additional shielding can be placed (see comment below). The average neutron flux 
at 90 degrees for all energies is 1.88 $\times 10^{-15} n/p/mm^{2}$.
Of greatest interest in this study is the flux at 90 degrees, or 
neutrons penetrating through the thinnest point of the bulk 
shielding. It should be noted from Fig.~\ref{fig:005} that there is an 
appreciable number of neutrons at 0 and 180 degrees that escape the
shielding block. The very high 'wings' at low and high angles point to holes 
in the shielding for beam entry and target servicing. As seen below, 
addition of more shielding both upstream and downstream can 
adequately control these higher fluxes. As distance along the axis 
of the beam does not impact cavern size, such additions have little 
consequence on the rock excavation question.

\begin{figure}[bh!] 
\centering
\includegraphics[width=0.95\textwidth]{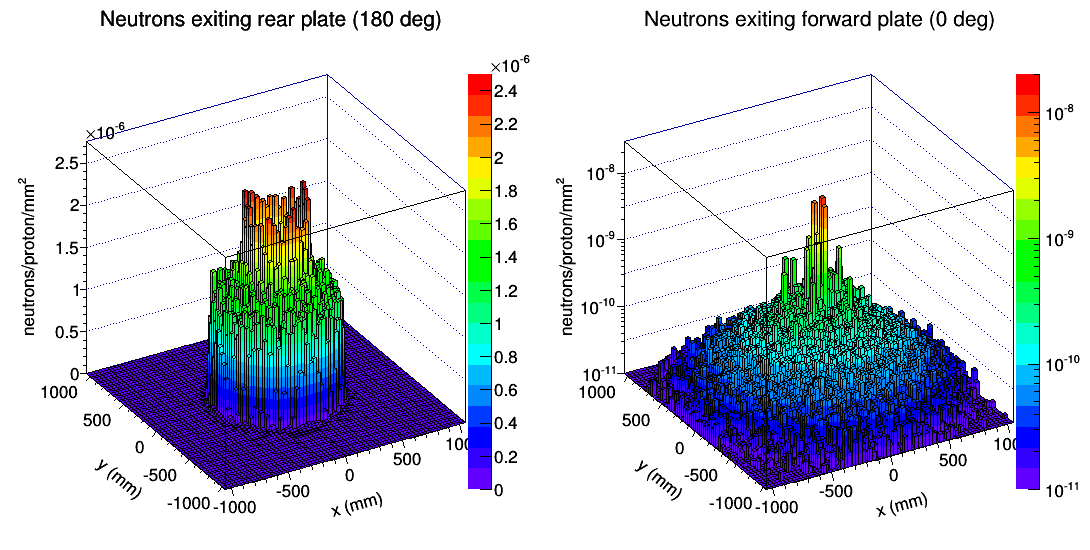}
\caption{Neutron flux recorded on detector plates placed at 0 and 180
  deg on the target shielding. The shielding is made of 
  30 cm steel and 90 cm boron rich concrete.}
\label{fig:006}
\end{figure}

\begin{figure}[h!] 
\centering
\includegraphics[width=0.6\textwidth]{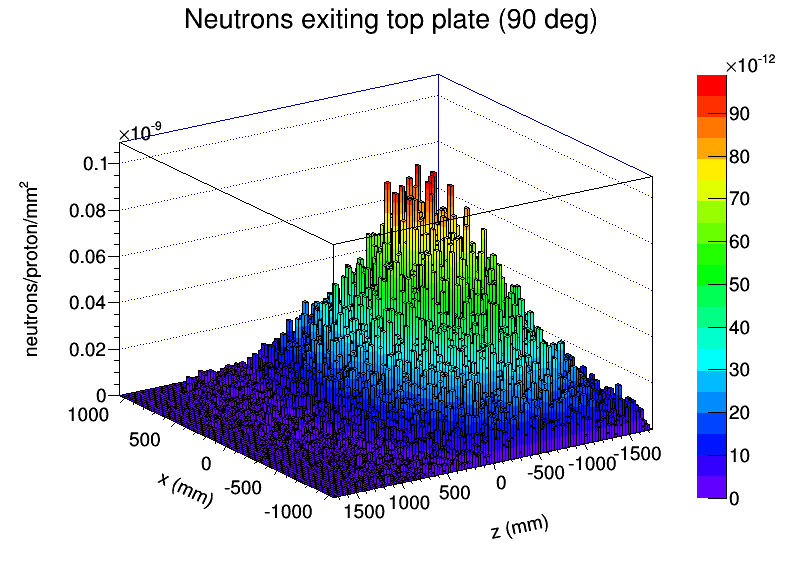}
\caption{Neutron flux at 90 deg for a shielding consisting of 30 cm
  steel and 90 cm boron rich concrete. The neutron levels can be
  translated into rock activation, as shown in the next section.}
\label{fig:007}
\end{figure}

To obtain a more accurate calculation of the neutron flux, simulated detector plates were placed on the
shielding block to record the flux at 0, 90 and 180 degrees for this baseline shielding configuration with 30 cm of steel and 90 cm of boron rich concrete. The neutron flux detected at the
front and at the back of the shielding block is shown in Fig.~\ref{fig:006}.
The neutron contamination in the proton beam pipe has a peak value of 2.4$\times
10^{-6} n/p/mm^{2}$ while the flux in the space left for the wobbler magnets has a peak
value of 1.4$\times 10^{-6} n/p/mm^{2}$. 
The neutron flux detected on
the plate above the shielding block (Fig.~\ref{fig:007}) is forward 
biased,  as most of the neutrons that are backscattered escape in the space left for the
magnets and therefore have no chance to be scattered into the
sleeve and detected on the first half of the plate. The average neutron
flux is 4$\times 10^{-11} n/p/mm^{2}$.

\vspace{0.5 cm}

\subsection{Design Optimisation}\label{sec:ooo}

\begin{figure}[hb!] 
\centering
\includegraphics[width=0.95\textwidth]{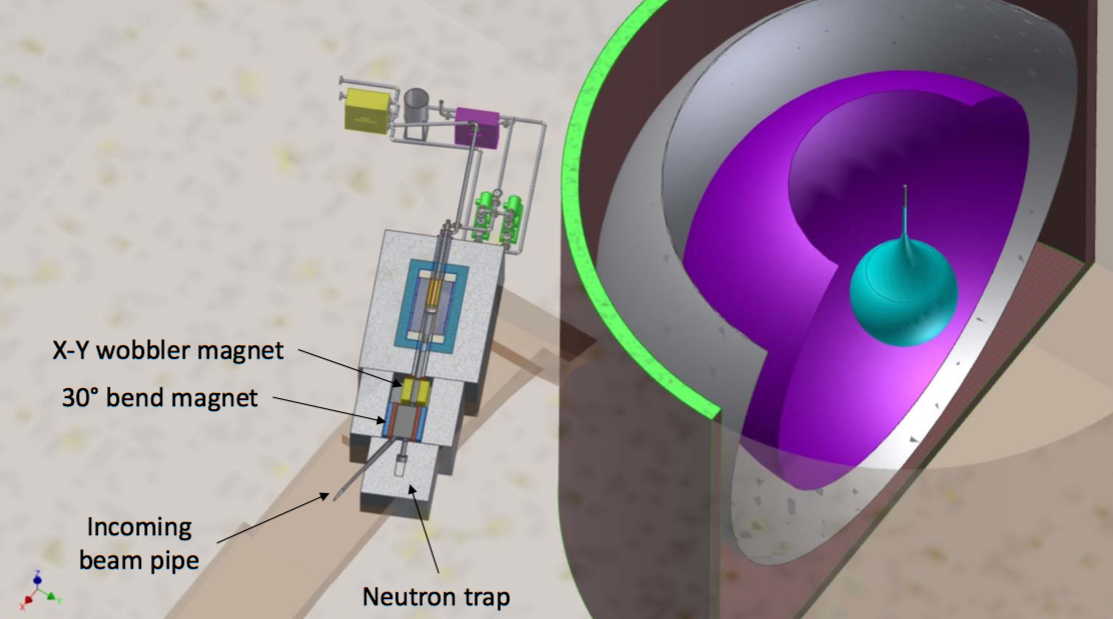}
\caption{Aerial view of the target and shielding system, with the wobbler magnet and bending magnet placed outside the shielding block. The KamLAND detector with the surrounding buffer region and water layer is also shown.  }
\label{fig:008}
\end{figure}

\noindent
Slight changes to this design were imposed by the fact that a large
fraction of backscattered neutrons escape in the space left for the 
wobbler magnets without being moderated and captured by
shielding. To avoid this, the hole at the front end of the target 
system was sealed with shielding, leaving space only for the proton 
beam pipe. The wobbler magnet will be placed outside the
shielding block, as well as the 30 degree bending magnet. 
This new configuration will adequately block the escaping neutron
flux in the upstream direction (Fig.~\ref{fig:008}). 
The forward neutron flux (at low angles) can be further attenuated 
by using movable blocks of steel and concrete. There is much less 
of a space restriction at the beam height, so added material can be 
readily provided. This extra material needs to be movable to enable 
changing of the target assemblies, which occurs from the downstream 
side of the target complex.  

\vspace{0.2 cm}

\begin{figure}[h!] 
\centering
\includegraphics[width=0.70\textwidth]{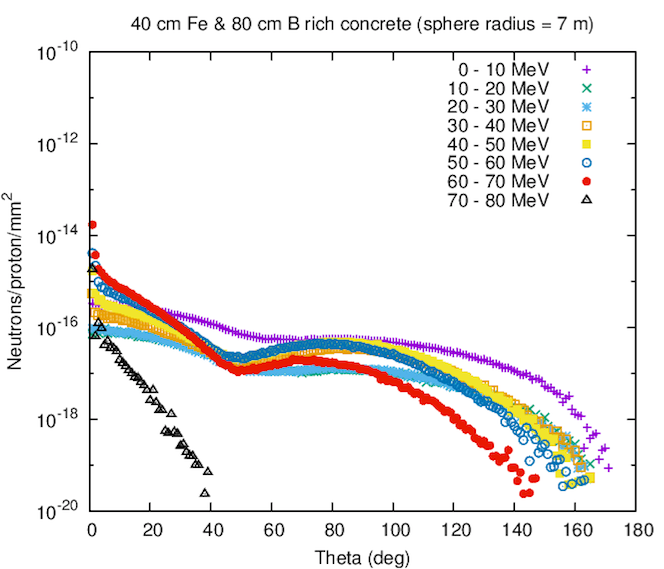}
\caption{The neutron flux detected on a sphere surrounding the
  shielding after seal in the space for magnets. The shielding material
  consists of 40 cm steel and 80 cm boron rich concrete.}
\label{fig:008a}
\end{figure}

\noindent
Figure~\ref{fig:008a} shows that the neutron flux after optimization
(the holes at the front and at the back of the target were filled with
shielding material and the new material combination was changed to 40 cm
steel and 80 cm boron rich concrete for better neutron attenuation). 
The neutron flux is much
lowered in the optimized design as seen in Fig.~\ref{fig:008a}. The 
detector sphere radius is 7 m as it surrounds not just the target system but 
also the wobbler magnet and the neutron trap.

\vspace{0.2 cm}

\section{Rock Activation Studies}\label{sec:ddd}
 
\noindent
Although the parametric studies of neutron flux versus shielding composition suggested an optimized preliminary design for which the neutron flux was below the desired value, the ultimate shielding effectiveness is given by the rock activation of cavern wall which must not exceed  0.1 Bq/g. Therefore, the calculation of induced activation 
 and the analysis of the radionuclides produced serve as the guide in designing 
the final shielding configuration. 


\vspace{0.2 cm}
\subsection{Methodology for Calculation of Induced Activity}\label{sec:hhh}
\vspace{0.2 cm}

The time dependence of the number of radioisotopes produced is given by
the individual production and decay rate of each isotope. The 
production rate for a radioactive isotope is given by:

\begin{equation}\label{eq:prod_rate}
    {dN_{i}^{prod}\over dt}={N_{iso}\over \Delta t}={{N_{iso} I }\over{N_{p}  e}}
\end{equation}

\noindent
where $N_{iso}$ is the number of isotopes produced by the simulation,
$I$ is the current, $N_{p}$ is the number of simulated protons, and $e$ is the proton charge. 
Once the radioisotopes are produced, they decay
exponentially with time. The decay rate for each isotope $i$ is a
function of its decay constant:

\begin{equation}\label{eq:decay_rate}
{{dN_{i}^{decay}(t) \over dt}}= -{\lambda_{i}N_{i}(t)}
\end{equation}

During the beam-on period, the time evolution 
can be obtained by combining the production and decay rates:

\begin{equation}\label{eq:rate}
    {{dN_{i}(t)\over dt}}={{N_{iso} I}\over{N_{p} e}}-{\lambda_{i}N_{i}(t)}
\end{equation}

The solution of Eq.~\ref{eq:rate}, gives the number of isotopes at any 
time $t$ during the beam exposure:

\begin{equation}\label{eq:iso_beamon}
    {N_{i}(t)}={{{N_{iso} I}\over{N_{p} e \lambda_{i}}}{(1-\exp(-\lambda_{i}t))}}
\end{equation}

However, after the beam is switched off, following a continuous 
exposure for a time $t_{1}$, the number of isotopes after a time 
$t$, which includes both the beam on period $t_{1}$ as well as a 
beam off period $t_{2}$, is given by:

\begin{equation}\label{eq:iso_beamoff}
    {N_{i}(t)}={{{N_{iso} I}\over{N_{p} e \lambda_{i}}}{(1-\exp(-\lambda_{i}t_{1}))\exp(-\lambda_{i}(t_{2}))}}
\end{equation}

 The induced activity given by one particular isotope is given by Eq.~\ref{eq:iso_activity}:

\begin{equation}\label{eq:iso_activity}
    {A_{i}(t)}={\lambda_{i}N_{i}(t)}
\end{equation}

Using Eq.~\ref{eq:iso_beamoff}  and Eq.~\ref{eq:iso_activity}, the activity of each isotope produced
can be given at any given time $t$ which includes the beam on time $t_{1}$ and the
beam off time $t_{2}$. The total induced activity given by all
isotopes is given by Eq.~\ref{eq:total_activity}:

\begin{equation}\label{eq:total_activity}
   {A(t)} = {\sum_{i}  {\lambda_{i}N_{i}(t)} }
\end{equation}

The production rates of all isotopes produced in the rock are required further for calculation of the total induced activity. 

\subsection{KamLAND Rock Composition}\label{sec:ooo}

KamLAND is located under the peak of Ikenoyama (Ike Mountain,
36.42$^{\circ}$N, 137.31$^{\circ}$E). Various types of rocks are found
in Ikenoyama in unknown quantities, such as Inishi type rocks, skarn
rocks, but also granite and limestone. The Inishi type rock is
characteristic for the Japanese mountains and is made of various
oxides with a high concentration of $SiO_{2}$. Skarns are 
calcium-bearing silicate rocks that are most often formed
at the contact zone between intrusions of granitic magma bodies and
carbonate sedimentary rocks such as limestone and dolostone. The
skarn-type rock is defined as a combination of 70\% granite and 30\%
limestone. The specific gravity for generic skarn is 2.75\% $g/cm^{3}$ and
for the Inishi rock is 2.65 \% $g/cm^{3}$~\cite{Abe}.  The exact composition
of the Inishi rock is given in Table~\ref{tab:Inishi-rock}~\cite{Tang}~\cite{Lindley}. 

 \begin{table}[h!]
\centering
\caption{\label{tab:Inishi-rock} Chemical composition of the
  Inishi-type rock in elemental percentage.}
\vspace{0.2cm}
\begin{tabular}{cccc}
\hline\hline
 Compound & Composition ($\%$ ) &Compound & Composition ($\%$ )\\
\hline
{$SiO_{2}$}  & 60.70 & CaO & 6.00 \\
{$TiO_{2}$}  & 0.31 & {$Na_{2}O$}& 6.42 \\
{$Al_{2}O_{3}$}  & 17.39 & {$K_{2}O$} & 3.47 \\
{$Fe_{2}O_{3}$} &  1.10 & {$P_{2}O_{5}$} & 0.18 \\
FeO & 1.22 & {$H_{2}O$} & 1.27 \\
MnO & 0.15 & S & 0.01 \\
MgO  & 0.93 & {$CO_{2}$} & 0.96 \\
\hline\hline
 \end{tabular}
 \end{table}

\subsection{List of Radionuclides Produced}\label{sec:ooo}

\begin{figure}[th!] 
\centering
\includegraphics[width=1.0\textwidth]{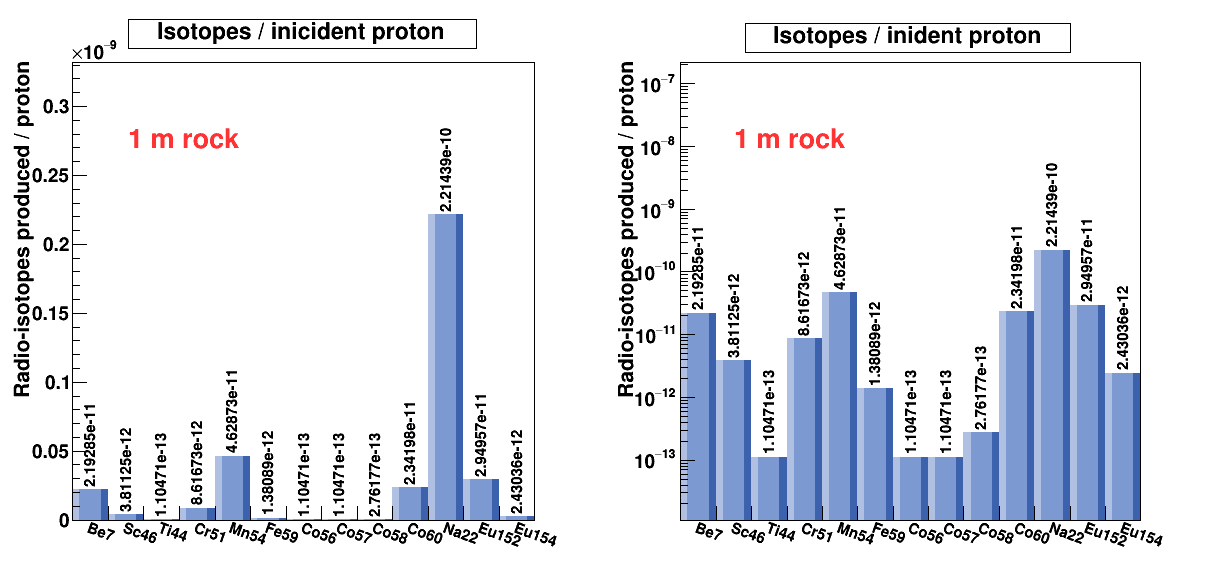}
\caption{Rate of the isotope production in 1 m of Inishi-type rock. The
  production rate is dominated by the long-lived isotopes. The shielding consists of 40 cm steel and 80 cm boron concrete. }
\label{fig:009}
\end{figure}

Composition analysis of rock samples collected in the mine and
irradiated in the MIT reactor indicated that the rocks contained traces
of cobalt and europium. The cobalt concentration in the rock samples
varied from 1 to 30 ppm by weight, while europium concentrations
averaged around 1 ppm. These element concentrations were taken into
account for the above 
Inishi-rock composition.  Geant4 simulations show that 
the isotopes that are produced in the rock are: $^{7}Be$, $^{46}Sc$, $^{44}Ti$,
$^{51}Cr$, $^{54}Mn$, $^{59}Fe$, $^{56}Co$, $^{57}Co$, $^{58}Co$,
$^{60}Co$, $^{22}Na$,  $^{152}Eu$, $^{154}Eu$, $^{134}Eu$ and $^{134}Cs$.
(Fig.~\ref{fig:009}).

Many of these isotopes were seen in the MIT
activation spectra. One notable not seen is  $^{22}Na$ which has a
13 MeV threshold due to the (n, 2n) production channel. No neutrons at
the reactor have this high energy, however the neutron spectrum from
the IsoDAR target does have this and higher energies. The majority of 
these isotopes have a half life of several days, but $^{60}Co$,
$^{22}Na$,  $^{152}Eu$, and $^{154}Eu$ dominate the production rate 
and  have a lifetime range between 2.6 - 13.5 years. Also, $^{44}Ti$ 
has a lifetime of 63 years, but this isotope was found in small
quantities. Table~\ref{tab:isotopes} shows the characteristics of the
isotopes of interest.

\vspace{0.2cm}
 \begin{table}[h!]
\centering
\caption{\label{tab:isotopes} Characteristics of the isotopes of interest.}
\vspace{0.2cm}
\begin{tabular}{cccccc}
\hline\hline
Isotope & Half life & Progenitor & Concentration& Neutron energy&Cross section(barns)\\
\hline
$^{60}Co$ & 5.2 y & $^{59}Co$ & ~30 ppm & Thermal & ~20\\
$^{152}Eu$ & 13.5 y & $^{151}Eu$ & ~1 ppm & Thermal & ~50000\\
$^{154}Eu$ & 8.6 y & $^{153}Eu$ & ~1 ppm & Thermal & ~2000\\
$^{22}Na$ & 2.6 y & $^{23}Na$(n,2n) & ~5\%  & >13 MeV & 0.015\\
\hline\hline
 \end{tabular}
 \end{table}
\vspace{0.2 cm}

\vspace{0.2 cm}
\subsection{Spatial Distribution of Activity Induced at Various Depths}\label{sec:ooo}

\vspace{0.2 cm}

\begin{figure}[h!] 
  \centering
  \includegraphics[width=0.9\textwidth]{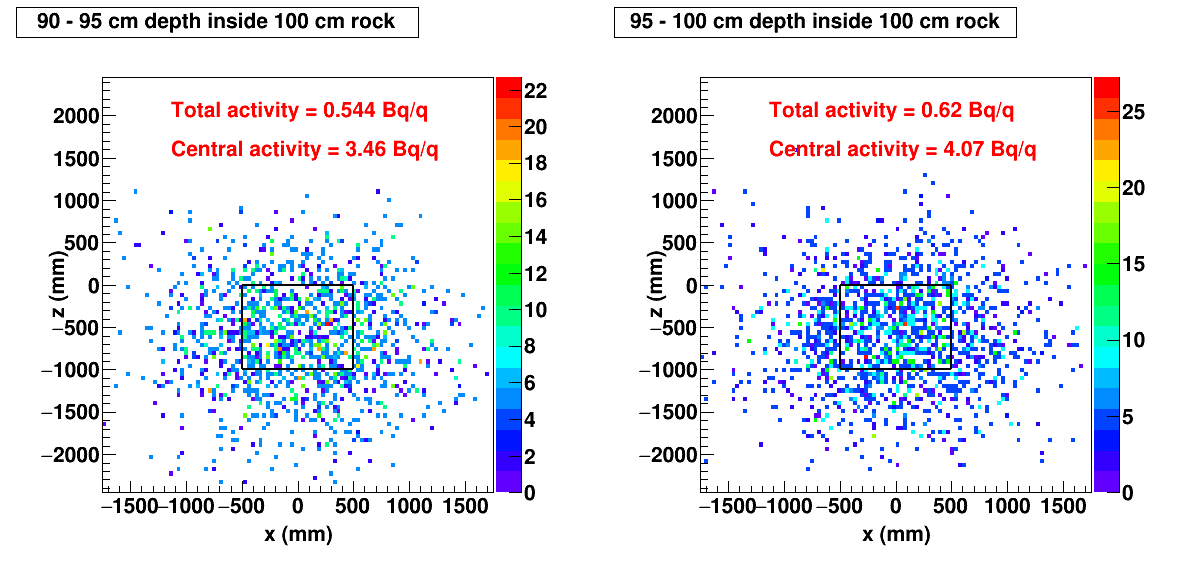}
\caption{Spatial distribution of induced activity in the lowest layers
  of the considered rock sample, in the close proximity of the target
  and shielding system (90-100 cm). The overall activity in the
  layers is above the required limit of 0.1 Bq/g and much higher in the central hotspots. This analysis is for the 40 cm steel and 80 cm boron concrete shielding.}
\label{fig:011}
\end{figure}

\begin{figure}[h!] 
\centering
\includegraphics[width=1.0\textwidth]{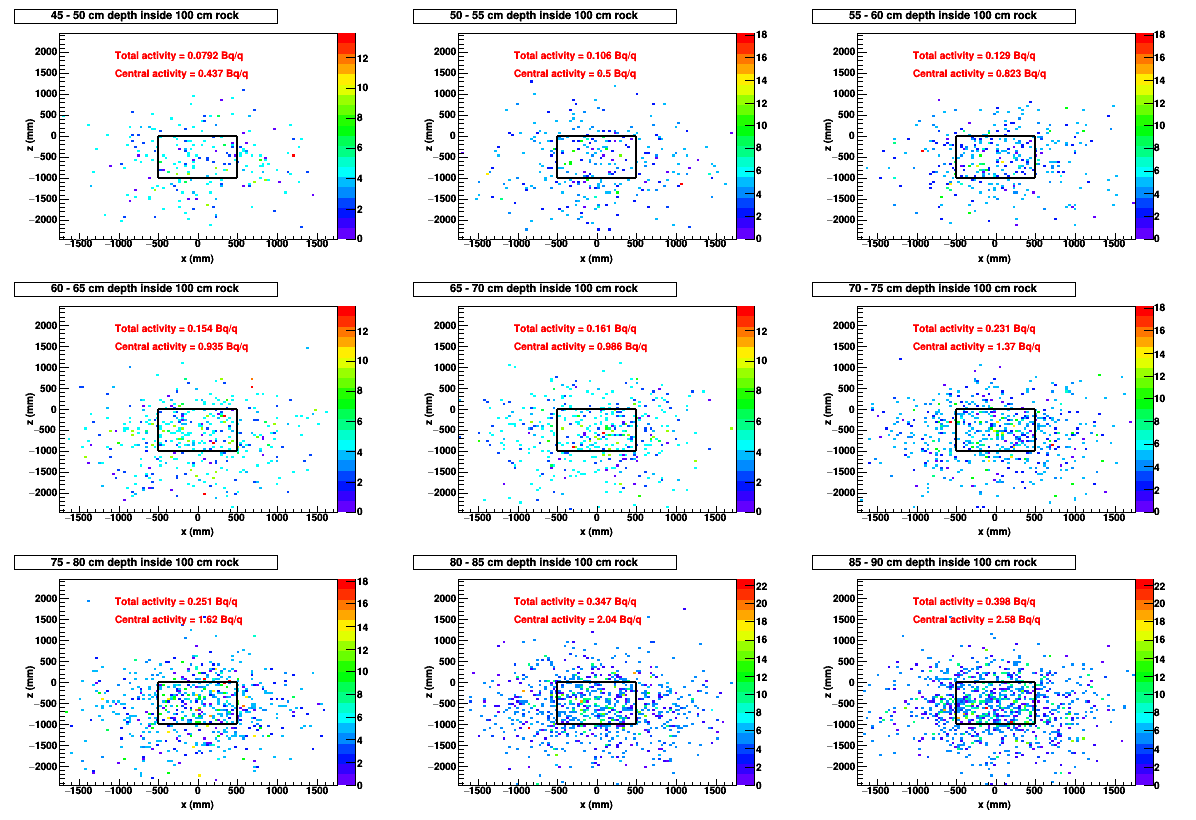}
\caption{Spatial distribution of induced activity in the upper layers
  of the considered rock sample (45-90 cm) for the 40 cm steel and 80 cm boron concrete shielding.The overall activity in the
  layers is above the required limit of 0.1 Bq/g. }
\label{fig:012}
\end{figure}

A rock sample having the same area  as the top part of the 
shielding block and a thickness of 100 cm, was considered in our 
simulations. The rock sample was placed on top of the
shielding block in the Monte Carlo modelling. The rock composition 
is the same as in Table~\ref{tab:Inishi-rock} with Co and Eu 
fractions by weight added. Using the equations above, one can calculate the spatial distribution
of induced activity of all isotopes produced at various
depths. The rock sample was sliced into 20 layers, each
having 5 cm thickness. The total activity is averaged over the entire
rock area, however there is a central hot spot in the rock sample where the
distribution is approximately uniform and where the activity is
averaged over the hot spot size  of 100$\times$100 cm. 
The activity is calculated after 5 years beam on and 2 years beam 
cool down time for the 40 cm steel and 80 cm boron concrete shielding. 
Figs.~\ref{fig:011} and ~\ref{fig:012} 
show the activity in the first 50 cm rock close to the target
shielding. The overall activity in the
rock is above the required level of 0.1 Bq/g, 
and much higher on the central hotspots. 
As most of the isotopes are short lived, the contribution to the total
activity after seven years must be associated with the long lived
isotopes like  $^{60}Co$, $^{152}Eu$,  $^{154}Eu$ and $^{22}Na$. The
contribution for these isotopes to the total activity in Bq/g in the central hotspots is shown in Table~\ref{tab:activity}
for all 20 rock layers.  The $^{22}Na$ gives the highest activation and
it is produced by the fast neutrons escaping from the shielding, 
indicating that thicker layers of steel are needed for extra safety. 

 \begin{table}[h!]
\centering
\caption{\label{tab:activity} Contribution to the total induced
  activity of all 13 isotopes in the central hot spots (100 $\times$ 
  100 cm) in Bq/g, given by the long-lived
  isotopes of interest.}
\vspace{0.2cm}
\begin{tabular}{cccccc}
\hline\hline
 Rock layer (cm)&  $^{60}Co$ &$^{152}Eu$ &$^{154}Eu$& $^{22}Na$&Total\\
\hline
0-5&0&0&0&0.014&0.053\\
5-10&0&0.011&0&0.034&0.052\\
10-15&0.006&0.011&0&0.061&0.069\\
15-20&0.017&0.021&0&0.007&0.125\\
20-25&0.040&0.005&0&0.061&0.109\\
25-30&0.040&0.021&0.005&0.109&0.151\\
30-35&0.063&0.021&0&0.143&0.208\\
35-40&0.046&0.043&0&0.232&0.238\\
40-45&0.069&0.048&0.005&0.348&0.356\\
45-50&0.063&0.043&0.005&0.286&0.437\\
50-55&0.155&0.048&0&0.354&0.500\\
55-60&0.115&0.064&0.005&0.423&0.823\\
60-65&0.126&0.101&0.016&0.682&0.935\\
65-70&0.143&0.069&0&0.791&0.986\\
70-75&0.138&0.080&0&1.070&1.370\\
75-80&0.092&0.080&0.011&1.360&1.620\\
80-85&0.109&0.112&0.005&1.830&2.040\\
85-90&0.149&0.085&0&1.960&2.580\\
90-95&0.080&0.053&0.011&2.840&3.460\\
95-100&0.063&0.016&0.005&3.550&4.070\\
\hline\hline
 \end{tabular}
 \end{table}

\newpage

\vspace{0.2cm}

\begin{figure}[h!] 
  \centering
  \includegraphics[width=0.6\textwidth]{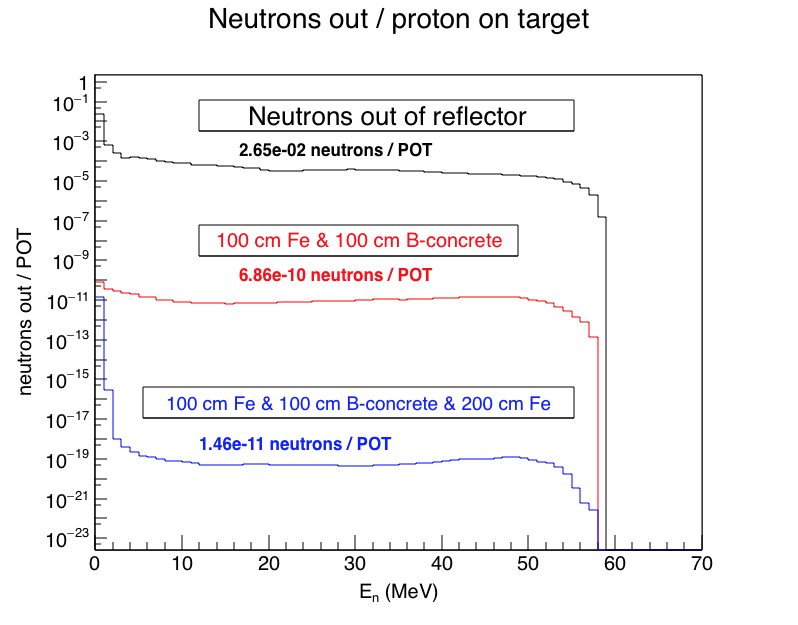}
\caption{The neutron spectra out of the reflector and out of the 2 m
  shielding. It also shows the neutron spectra for the 4 m shielding side that fills the
  available space between the target system and the KamLAND detector (Fig.~\ref{fig:008}). }
\label{fig:013}
\end{figure}
\begin{figure}[h!] 
  \centering
  \includegraphics[width=0.87\textwidth]{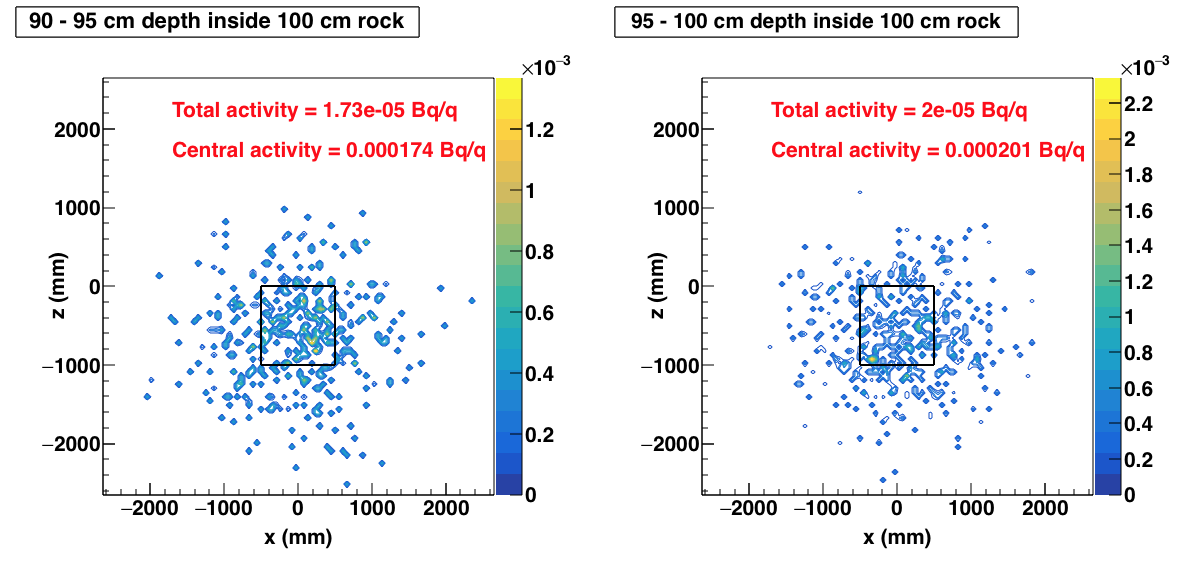}
\caption{Spatial distribution of induced activity in the lowest layers
  of the considered rock sample, in the close proximity of the target
  and shielding system (90-100 cm). The overall activity in the
  layers is below the required limit of 0.1 Bq/g even in the central 
  hotspots. This analysis is for the 200 cm shielding.}
\label{fig:014}
\end{figure}

As rock activation studies ultimately evaluate the effectiveness of 
the shielding, a 2 m shielding consisting of 1 m of steel and 1 m 
of boron concrete was considered for calculations of the induced 
activation in the rock layers. The neutron spectra out of the
reflector and out of the 2 m shielding is shown in
Fig.~\ref{fig:013}. This plot also shows the neutron spectra out of
the 4 m shielding side towards the KamLAND detector that will be
placed between the target and the detector to suppress the neutron and
gamma background (Fig.~\ref{fig:008}). The 2 m shielding decreases the
neutron rates from $2.65\times10^{-2} $neutrons/POT (out of the
reflector) to $6.86\times10^{-10} $neutrons/POT. With this extra shielding,  the 
activation levels on central spots in the rock layers will also decrease
significantly as Fig.~\ref{fig:014} shows for the first 10 cm of
rock situated in the proximity of the shielding.

The total activity and the activity on the central hotspots for each layer of rock sample was calculated for the 200 cm shielding and the results are shown in Table~\ref{tab:activity2}.

 \begin{table}[h!]
\centering
\caption{\label{tab:activity2} The total and central activity on
  hotspots in each layer of the rock}
\vspace{0.2cm}
\begin{tabular}{ccc}
\hline\hline
 Rock layer (cm)&  Total activity: $10^{-6}(Bq/g)$& Central activity: $ 10^{-5}(Bq/g)$\\
\hline
0-5&0.116&0.113\\
5-10&0.407&0.113\\
10-15&0.663&0.175\\
15-20&0.856&0.917\\
20-25&0.84&1.04\\
25-30&0.859&0.661\\
30-35&1.10&1.07\\
35-40&1.48&1.02\\
40-45&2.4&1.42\\
45-50&2.5&2.53\\
50-55&3.25&2.97\\
55-60&3.89&3.36\\
60-65&5.01&4.3\\
65-70&5.87&4.53\\
70-75&8.28&6.54\\
75-80&8.94&9.4\\
80-85&12.5&10.9\\
85-90&13.4&10.5\\
90-95&17.3&17.4\\
95-100&20&20.1\\
\hline\hline
 \end{tabular}
 \end{table}

\subsection{Activity induced in the rock}\label{sec:ooo}

\begin{figure}[h!]  
\centering
\includegraphics[width=0.75\textwidth]{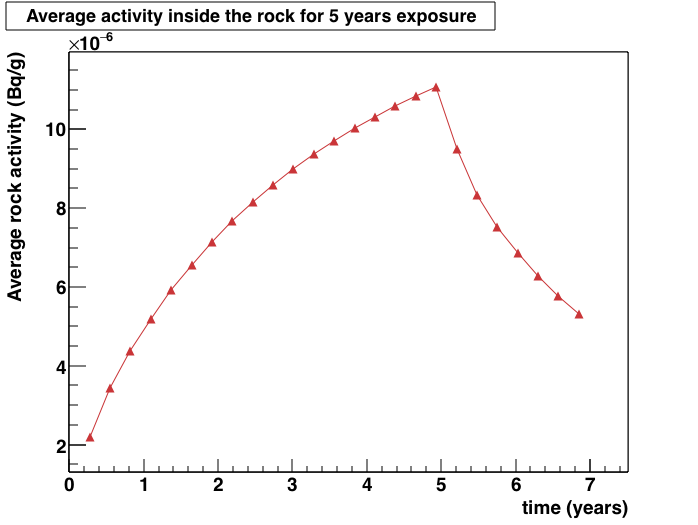}
\caption{The total averaged activity in the entire rock during beam On
and beam Off time. The total shielding thickness is 200 cm.}
\label{fig:activity}
\end{figure}

The total activity in the entire rock sample considered after 5 years
run and 2 years beam cool down period is shown in  Fig.~\ref{fig:activity} with
the peak value representing the activity at the beam switch off
time. It can be seen that, for the 200 cm of  shielding materials, the induced activity inside the rock is well below the imposed limit of 0.1 Bq/g, both throughout the beam on period, as well as after the beam switch off.

\subsection{Distribution of Isotope Production with Rock Height}\label{sec:ooo}

At the rock surface the slow neutrons will be responsible for 
surface activation, but faster neutrons  will also produce 
activation at higher levels inside the rock. The isotope production inside the rock varies with the distance from the top layer of shielding. 
A large fraction of isotopes are produced in
close proximity of target-shielding system with the production 
decreasing with the rock height and reaching a minimal value after 1 m of rock. 

\begin{figure}[h!] 
\centering
\includegraphics[width=0.8\textwidth]{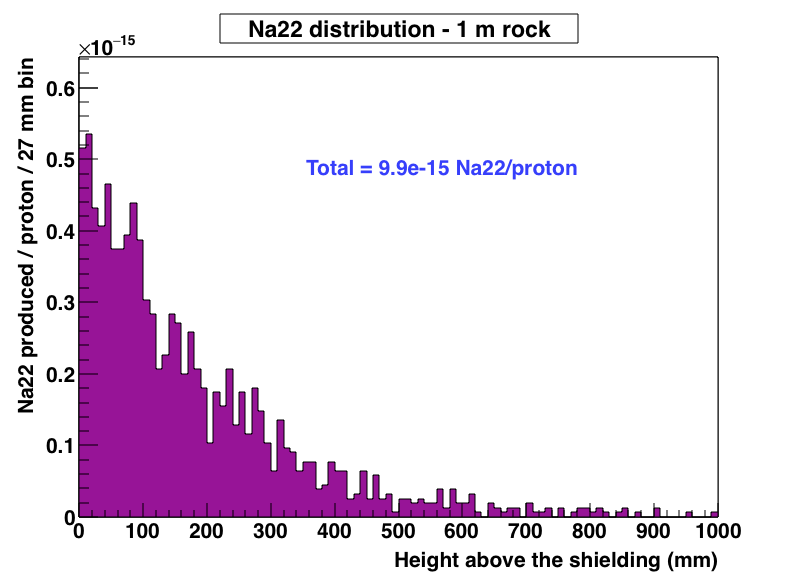}
\caption{ $^{22}Na$ production as a function of rock height for the 200 cm shielding (100 cm steel and 100 cm boron concrete).}
\label{fig:010}
\end{figure}

The $^{22}Na$ production as a function of rock 
height is shown in Fig.~\ref{fig:010}.
As $^{22}Na$  is a significant component of the rock chemical composition,
even a small flux of high energy neutrons can contribute to an
unacceptable high $^{22}Na$ production.

\section{Shielding Requirements for Reducing the Neutron and Photon Physics Backgrounds in the KamLAND Detector}

While the majority of non-beam backgrounds for IsoDAR at KamLAND can be experimentally measured and subtracted by comparing beam-on versus beam-off measurement periods, there are a few backgrounds inherent to the beam and target itself that must be accounted for. Most significant of these are neutrons produced by the interaction of the proton beam on the Li-Be target. Many of these neutrons will be captured on the $^7$Li, which will then decay to produce the electron antineutrinos that are to be measured in the experiment, but there is a significant portion that will not. Geant4 simulations find that approximately $2.65 \times 10^{-2}$ neutrons per proton-on-target (POT) will escape the neutrino-producing reflector and enter the target shielding. Due to the calculated $7.88 \times 10^{24}$ POT over IsoDAR's five year run, this represents an unacceptably high background of neutrons into the KamLAND detector that the needs to be reduced through shielding.

IsoDAR will study two kinds of neutrino events during its run. The first are inverse beta decay (IBD) events, of which there will be $8.2 \times 10^{5}$ events over the five year run. These events are characterized by a prompt positron signal with $E_{vis} = E_{\bar{\nu}_e} - 0.78$ MeV coincident with a delayed neutron capture releasing a 2.2 MeV gamma within ~200 $\mu$s. These events will not be affected by the neutron background since the signal is a two-part delayed coincidence. The second kind of event to be studied is the sample of low-energy $\bar{\nu}_e$-electron scatters (ES). There will be approximately 2600 of these events above a 3 MeV threshold over the five year run, and they will be easily mimicked by elastic scatters of neutrons in the KamLAND detector. Thus, additional shielding is needed to reduce the rate of neutrons and gammas over 3 MeV in the KamLAND detector region to a rate less than about 200 over the five year run.

\subsection{Neutron Shielding}

\subsubsection{Target Shielding}

The IsoDAR target is already shielded by 200 cm of material intended to absorb neutrons in order to comply with Japan's radiation requirements for the surrounding cavern. A full Geant4 simulation from the incoming POT to the neutrons escaping from this shielding showed a final output of just $6.86 \times 10^{-10}$ neutrons per POT, which drops to $5.40 \times 10^{-10}$ above the 3 MeV threshold. The majority of these neutrons are oriented away from the KamLAND detector due to the geometry of the target shielding, with many escaping through the vacuum of the beam pipe or the less heavily shielded water access tubes. Additional shielding is placed outside the target cube to further shield these accesses in the interests of radiation reduction, and also helpfully prevents the majority of these neutrons from moving towards the detector. 

After the target shielding and accounting for the limited solid angle intercepted by the KamLAND detector, the incoming neutron flux towards the KamLAND detector is reduced to $1.069 \times 10^{-9}$ neutrons above 3 MeV per POT. This estimate is made after a full simulation starting from incoming beam protons at 60 MeV through the entire target geometry, and then counting the neutrons whose direction of motion will intercept the detector itself, with the assumption that neutrons will scatter as much out of the intercepting solid angle as into it. This assumption allows us to use straight line trajectories for neutrons leaving the target region for the purposes of estimating neutron flux in the detector.

\vspace{0.5 cm}
\subsubsection{Additional Shielding Design}
\vspace{0.5 cm}

\begin{figure}[ht!]
\begin{center}
\includegraphics[width=0.75\textwidth]
{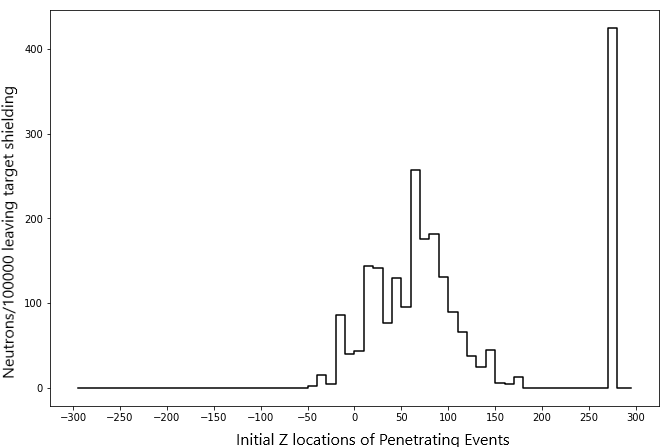}
\caption{\label{Zlocations} The initial z (along the beam direction with positive downstream) locations of a sample of the neutron events which successfully passed through the KamLAND shielding during the first round of simulation (without the additional 2m of iron). The large spike on the far right represents the excess of neutrons escaping the shielding through the water cooling pipes, which shall be countered by a downstream shift of the additional shielding.} 
\end{center}
\end{figure}

The additional shielding will be placed between the IsoDAR target and the KamLAND detector complex, and will be composed of stacked blocks of steel to produce a solid rectangle 2 m thick, 5 m high, and 5 m wide. This shielding block will be slightly off center from the target, with center around 50 cm downstream from the center of IsoDAR. This is because simulations of the full IsoDAR target from POT to escaping neutrons show that the majority of high energy neutrons, that the shielding block is meant to moderate, escape on the downstream end of the target (See Fig.~\ref{Zlocations}). 

This is due to the water cooling pipes on that end, which provide neutrons an easier path out of the shielding. Unlike the beam pipe on the upstream end, which provides a similar function, these pipes are not encased in additional shielding blocks, so a shift of the additional shielding will be necessary to prevent an excess of neutrons from that end. Fortunately, there is a demonstrated lack of penetrating events from the upstream end which enables this shift without increasing the neutron background by any significant amount.

\subsubsection{KamLAND Buffer Region and Additional Shielding}

Further simulations were based on the neutron energy and direction spectra out of the target shielding and focused on the effectiveness of the KamLAND buffer region in reducing the incoming neutron flux. Unfortunately, while the KamLAND buffer region, at minimum 100 cm of water and 250 cm of paraffin oil, is very effective at capturing and moderating lower energy neutrons less than 10 MeV, it is comparatively far less effective on higher energy neutrons greater than 30 MeV. Incoming neutrons with energy greater than 50 MeV would enter the detector with energy above the threshold of 3 MeV at rates above $1 \times 10^{-3}$ per neutron, i.e. for every neutron above 50 MeV entering the KamLAND buffer region there will be at least $1 \times 10^{-3}$ neutrons above 3 MeV entering the detector. This resulted in background rates several orders of magnitude above that desired, with a total above 3 MeV of $5.92 \times 10^{-17}$ neutrons per POT. Given the $7.88 \times 10^{24}$ POT over the 5 year run this yields a total background of $4.66 \times 10^{8}$ neutrons.

\begin{figure}[ht!]
\begin{center}
\includegraphics[width=1.0\textwidth]
{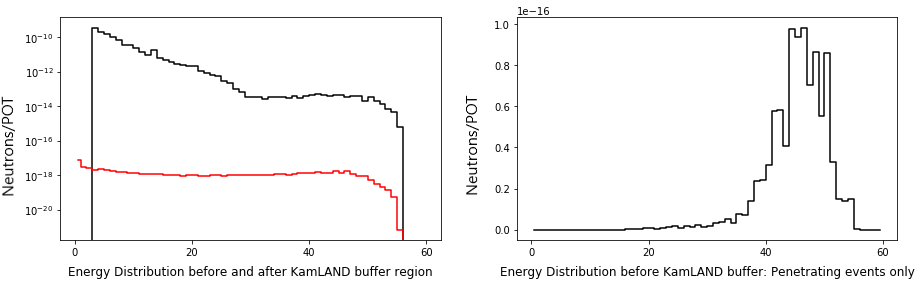}
\caption{\label{fig-noextrashielding} The above left shows the neutron energy distributions (in MeV) before (black) and after (red) the KamLAND buffer region without any additional shielding. The right plot shows the energy distribution of the penetrating neutrons only, i.e. the neutrons that reach the detector with energy $>3$ MeV.} 
\end{center}
\end{figure}

These studies demonstrated the need for additional shielding to be added in the 2 m space between the edge of KamLAND and the outer layer of the target shielding. Further, they allowed the prioritization of moderating or slowing high energy neutrons rather than all neutrons in this shielding, as seen in Fig.~\ref{fig-noextrashielding}. As different materials have different moderating effects on different energies of neutrons, this was very useful. Iron was selected as the material of choice in this shielding layer due to its high effective cross section for neutrons between 20 and 60 MeV, the energy domain for penetrating neutrons.

\begin{figure}[ht!]
\begin{center}
\includegraphics[width=0.75\textwidth]
{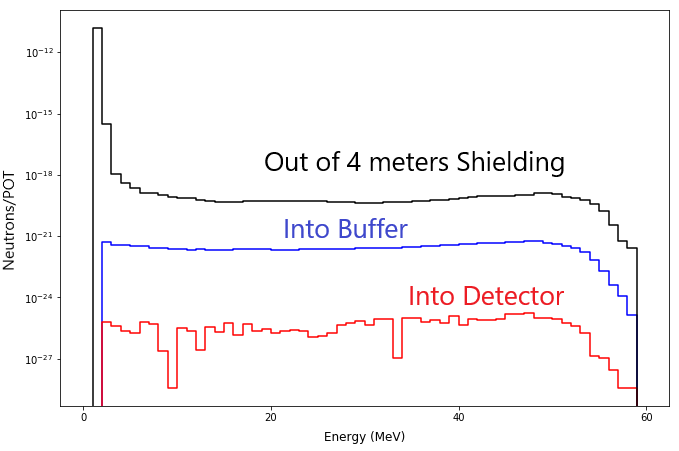}
\caption{\label{fig-fullSim} The displayed neutron energy distributions are taken from the simulation of the full geometry of the KamLAND detector and buffer. The simulation started with the neutron energy distribution out of the target and additional 2 m shielding (giving 4m total shielding) (black) and ended with the neutron energy distribution entering into the detector (red), with an intermediate energy check at the transition between the water and mineral oil buffer layers (blue).} 
\end{center}
\end{figure}

As demonstrated in Fig.~\ref{fig-fullSim}, the addition of 2m of iron shielding (bringing the total to 4 m) between the end of the target shielding and the start of the KamLAND buffer region drops the penetrating neutrons by nearly 8 orders of magnitude, especially in the higher energy domains where the buffer region proved insufficient. All combined, this total shielding left an estimated $2.78 \times 10^{-24}$ neutrons per POT above the 3 MeV threshold in the KamLAND detector region, which is sufficient to meet our requirements. This represents a mere 22 neutrons above 3 MeV in KamLAND over the entire 5 year run.

This estimate is further taken to be conservative due to the overestimation inherent in the two part simulation, which was made with a rounded up energy distribution to start the second part. IsoDAR will also benefit from beam timing checks that can be used to reduce slow neutron plus energy analyses of the final events. The final neutron distribution in the detector region is essentially homogenous from 3 to 55 MeV, while the ES signal events peak around 10 MeV and do not exceed 20 MeV, as this is the maximum neutrino energy from the $^8Li$ decays.

\subsection{Photon background}\label{sec:ffa}

\begin{figure}[h!]  
\centering
\includegraphics[width=0.6\textwidth]{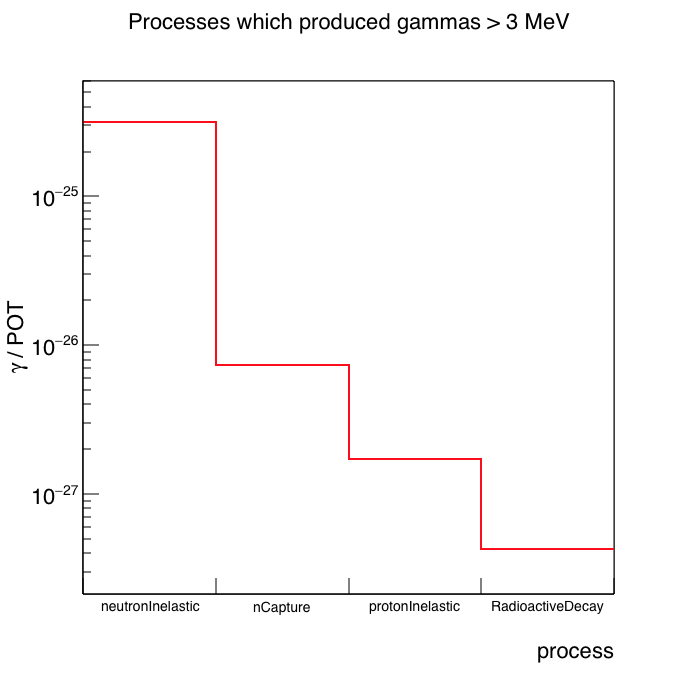}
\caption{The gammas with energies above 3 MeV that enter the Kamland detector are produced in the buffer regions mainly by neutron inelastic processes. Other processes like neutron capture, proton inelastic and radioactive decay have a lower contribution.}
\label{fig:entering}
\end{figure}

\begin{figure}[h!b]  
\centering
\includegraphics[width=0.7\textwidth]{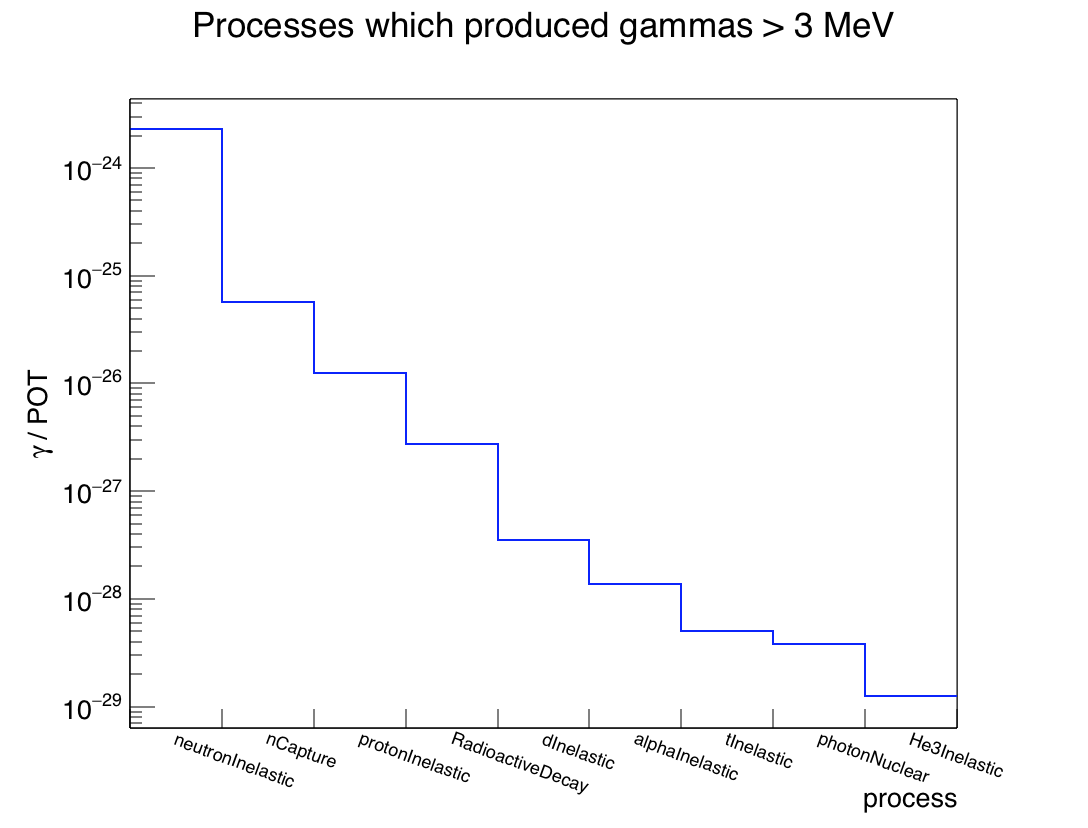}
\caption{Rate of processes that produce gammas with energies above 3 MeV for various particle interactions inside the Kamland detector. Apart from neutron inelastic, neutron capture, proton inelastic and radioactive decay, other processes that give a measurable contribution are deuteron, alpha, triton and He3 inelastic processes, and photon nuclear processes.}
\label{fig:produced}
\end{figure}

The same steel shielding also suppresses the gamma background towards KamLAND. As stated before, the
shielding was designed asymmetrically, having a larger thickness towards the detector (4 m)
with the extra 2 m filling the space between target and detector. 
For detector background studies only the neutrons having an energy
above 3 MeV are of concern since ES signal to be detected is above 3 MeV. 
The total number of gammas with energies above
3 MeV that enter the KamLAND detector in a solid angle of 0.17$\pi$ is $3.26\times10^{-25}$ gammas/POT.
These gammas are produced mainly by neutron inelastic processes 
($3.17\times10^{-25}$gammas/POT) but at smaller rates also by
neutron capture ($7.36\times10^{-27}$gammas/POT). (See Fig.~\ref{fig:entering}). Photon inelastic processes and
radioactive decay give a less significant contribution to the total
number of gammas. Low energy gammas are produced also in proton
inelastic, deuteron inelastic, ion inelastic, alpha inelastic and
photon nuclear processes, apart from the ones mentioned above (Fig.~\ref{fig:produced}).

The gamma background is calculated to be $2.38\times10^{-24}$ gammas/POT with energies above 3 MeV
and is produced in the detector mainly by inelastic interactions on
carbon ( $2.31\times10^{-24}$ gammas/POT ) and at a smaller rate by
neutron capture on hydrogen ($5.69\times10^{-26}$   gammas/POT) with a
peak energy of 2.2 MeV.  For 4 m of shielding towards the detector, the 
total numbers of gammas above 3 MeV is $2.7\times10^{-24}$ gammas/POT. 
Therefore, for $7.88\times10^{24}$protons/5 years run, this corresponds
to $\approx$ 22 gammas above 3 MeV for the 5 years experiment.

\section{Conclusion}\label{sec:ooo}

A shielding system for the IsoDAR neutrino experiment was
designed to meet the neutron flux irradiation requirements out of the shielding with a limiting value of 
$10^{-13} n/p/mm^{2}$. This value was obtained from neutron
irradiation of the Kamioka rock samples and analysis of the
radionuclides that were produced. New materials developed at Jefferson
Laboratory like boron concrete together with layers of steel were used to design the shielding in the confined space of
the current location in the mine, without significant rock
excavation. The radionuclides produced in the rock were identified and
the ones that give a significant contribution to the total induced activation
were the long lived isotopes like  $^{60}Co$,
$^{22}Na$,  $^{152}Eu$, and $^{154}Eu$. A spatial distribution of the
activity on the cavern wall identified the hot spots and the activity
for these spots is below the required limitation of 0.1 Bq/g after 5 years run plus 2 years cool down period for a 2 m  shielding. The neutron and gamma physics backgrounds in
the KamLAND detector were simulated for the 5 years experiment and it
was found that the levels at which they are produced are not a
significant background for detecting the IBD and ES physics signals.

\section*{Acknowledgements}

Adriana Bungau, Jose Alonso and Janet Conrad are supported by NSF PHY 1912764 and   
Larry Bartoszek, Edward Dunton and Michael Shaevitz are supported by NSF PHY 1707969. We thank Susan Kayser for the editorial support of this paper.

\end{document}